# Imaging real-space flat band localization in kagome magnet FeSn

**Authors:** Daniel Multer[1], Jia-Xin Yin[2]*, Md. Shafayat Hossain[1], Xian Yang[1], Brian C Sales[3], Hu Miao[3], William R Meier[4], Yu-Xiao Jiang[1], Yaofeng Xie[5], Pengcheng Dai[5], Jianpeng Liu[6], Hanbin Deng[2], Hechang Lei[7], Biao Lian[1]*, M. Zahid Hasan[1,8,9,10]*

**Affiliations:**

[1]Department of Physics, Princeton University, Princeton, 08544 New Jersey, USA.

[2]Department of Physics, Southern University of Science and Technology, 518055 Shenzhen, Guangdong, China.

[3]Materials Science and Technology Division, Oak Ridge National Laboratory, Oak Ridge, 37830 Tennessee, USA.

[4]Materials Science & Engineering Department, University of Tennessee Knoxville, Knoxville, 37921 Tennessee, USA.

[5]Department of Physics and Astronomy, Rice Center for Quantum Materials, Rice University, Houston, 77005 Texas, USA.

[6]School of Physical Science and Technology, ShanghaiTech University, 201210 Shanghai, China.

[7]Department of Physics and Beijing Key Laboratory of Opto-electronic Functional Materials & Micro-nano Devices, Renmin University of China, 100086 Beijing, China.

[8]Lawrence Berkeley National Laboratory, Berkeley, 94720 California, USA.

[9]Princeton Institute for the Science and Technology of Materials, Princeton University, Princeton, 08544 New Jersey, USA

[10]Quantum Science Center, Oak Ridge, 37830 Tennessee, USA

†Corresponding authors, E-mail yinjx@sustech.edu.cn; biao@princeton.edu; mzhasan@princeton.edu

**Kagome lattices host flat bands due to their frustrated lattice geometry, which leads to destructive quantum interference of electron wave functions. Here, we report imaging of the kagome flat band localization in real-space using scanning tunneling microscopy. We identify both the $Fe_3Sn$ kagome lattice layer and the $Sn_2$ honeycomb layer with atomic resolution in kagome antiferromagnet FeSn. On the $Fe_3Sn$ lattice, at the flat band energy determined by the angle resolved photoemission spectroscopy, tunneling spectroscopy detects an unusual state localized uniquely at the Fe kagome lattice network. We further show that the vectorial in-plane magnetic field manipulates the spatial anisotropy of the localization state within each kagome unit cell. Our results are consistent with the real-space flat band localization in the magnetic kagome lattice. We further discuss the magnetic tuning of flat band localization under the spin-orbit coupled magnetic kagome lattice model.**

# Introduction

Flat band occurs in quantum systems with localized wave functions, such as the Landau level[1,2], heavy fermion compounds[3], geometrically frustrated Lieb[4] and kagome lattices[5,6] and twisted bilayer graphene[7,8] *etc*. The research of flat band induced quantum effects can be traced back to the fractional quantum Hall effect[1,2]. The Landau levels generated by the application of a large magnetic field are perfect electronic flat bands, and a fractional filling of these flat bands can produce the fractional quantum Hall effect with anyonic excitations. Recently, kagome flat band has been widely discussed in several tantalizing quantum materials[9-19], exhibiting emergent physics of correlation and topology. However, in these kagome materials, a direct imaging of the flat band localization in real-space as well as its quantum tuning has been lacking.

In this work, we provide a rare example of tunable flat band localization in kagome magnet FeSn. Utilizing high-resolution scanning tunneling microscopy, we observe real-space localization features of the kagome flat band. Through vectorial magnetic field tuning, we further demonstrate the tunability of the flat band localization. Our results advance our microscopic understanding of the flat band localization in a kagome lattice.

## Results and discussion

FeSn has a hexagonal structure (space group P6/mmm) with lattice constants[20,21] $a = 5.3$ Å and $c = 4.5$ Å. It consists of a $Fe_3Sn$ kagome layer and a $Sn_2$ honeycomb layer with alternating stacking Fig. 1**a**. It orders antiferromagnetically below $T_N = 365$ K, where the Fe spin moments align ferromagnetically within a single kagome plane but antiferromagnetically from those in the neighboring kagome planes[22-26]. Kagome flat band (illustrated in Fig. 1**b**) in FeSn has been evidenced by photoemission data for bulk crystals[15,27] and planar tunneling data of few-layer thin films[28], which is around -0.2eV and arises from the $Fe_3Sn$ layer. Consistent with previous scanning tunneling microscopy study in CoSn family compounds[14,29-31], our cryogenic cleaving of FeSn often yields atomic steps with a unit-cell height ~4.5 Å, either of $Fe_3Sn$ terminations (Fig. 1**c** for an example) or $Sn_2$ terminations. Occasionally, we find that $Fe_3Sn$ termination and $Sn_2$ termination meet at a half-unit cell step edge (Fig. 1**d**). As the bonding distance in $Fe_3Sn$ layer is much shorter than that of $Sn_2$, it is more energetically favorable for the $Sn_2$ layers to be broken during the cleaving process. Based on this fact, there are two features that could help us to identify these two terminations. First, when $Sn_2$ and $Fe_3Sn$ surface meet at an atomic step edge, the up surface with a step edge should be $Sn_2$ and the lower integrate surface should be $Fe_3Sn$. Second, $Sn_2$ surface should favor (Sn) vacancy impurities while the $Fe_3Sn$ surface should favor (Sn) adatom impurities. These two features collectively help us to identify the two surfaces in Fig. 1**d**. We further resolve both the $Sn_2$ and $Fe_3Sn$ lattice layers with atomic resolution as shown in Figs. 1**e** and **f**, respectively. The $Sn_2$ surface clearly exhibit a honeycomb lattice, while we could not resolve further atomic detail of the $Fe_3Sn$ kagome lattice beyond its clear hexagonal lattice symmetry similar to other studies[9,30]. The atomically identified kagome lattice offers us a unique platform to study the kagome flat band.

**Imaging flat band localization**

Our angle-revolved photoemission data Fig. 1**g** as well as its curvature (similar to second derivative method) analysis (Fig. 1**h**) along a high-symmetry path confirms the existence of flat band near -0.2eV, similar to the existing studies[15,27,28]. The flatness of the -0.2eV band is reflected in its band width, which is 1 ~ 2 orders of magnitude smaller than other dispersive bands in FeSn. Tunneling spectrum at such high binding energy, however, often combines the local density of states and inelastic tunneling signals, particularly for correlated materials that are rich of bosonic excitations. The tunneling spectrums for $Fe_3Sn$ and $Sn_2$ lattice layers are displayed in Fig. 1**i**. The flat band state is expected to be mainly from the kagome lattice layer. This expectation is partially consistent with our observation the there is a broad bump near -0.2eV for the spectrum taken on the $Fe_3Sn$ lattice layer, while no apparent bump is detected near -0.2eV for the $Sn_2$ honeycomb lattice. We subtract the two spectrums to highlight this difference as in the inset of Fig. 1**i**. We have been aware that in the planar junction tunneling of few-layer thin films of FeSn (Schottky heterointerface of FeSn and an n-type semiconductor Nb-doped SrTiO3), a more pronounced flat band peak near -0.2eV is directly detected in the spectrum[28]. However, we have checked three different crystals with three different Pt/Ir tips, and our vacuum tunneling on the Fe₃Sn termination does not show a pronounced peak feature near -0.2eV. The reason for this difference remains a mystery for us, which might be related with the surface nature of the flat band (that contributes more to the tunneling spectrum for thin film samples) as well as the different tunneling matrix effects in these two different tunneling experiments.

Focusing on the Fe₃Sn atomic layer, we take spectrums on the Fe atom (that forms kagome lattice) and Sn atom, respectively (Fig. 2**a**). Their spectral difference also uncovers a pronounced state at -0.2eV (Fig. 2**b**). For scanning tunneling experiments, as the tip atoms inevitably convolute the local density of states of probing atomic lattices, the detection of such a spectral difference at the flat band energy is remarkable, and hints for spatial localization. dI/dV map at -0.2eV (Fig. 2**d**) taken on the Fe₃Sn kagome lattice (Fig. 2**c**) consistently reveal that the corresponding electronic states are localized to the kagome lattice sites. We have checked that with a different tunneling junction set-up as V = -0.1eV and I = 0.1nA, we obtain a similar dI/dV map pattern at V = -0.2eV with states localized to the kagome lattice sites, suggesting that the localization phenomenon does not arise from a junction set-up point effect. dI/dV maps taken at other energies within -0.3eV to +0.3eV (Fig. 2**e**) do not show such signatures of localized states or show a weaker localization of the kagome lattice (such as the map at energy -0.15V), confirming that the strong electronic localization to kagome lattice is uniquely detected at -0.2eV and agreeing well with our photoemission results. These spectroscopic data taken together support the realization of electron localization of kagome flat band state at -0.2eV. We also note that from data in Figs. 2**a** and **b**, the localization phenomenon belongs to a second-order tunneling feature in our data, while the first-order tunneling background can come from the states of other dispersive bands (intrinsic) and the finite spatial resolution of the tip leading to a spatial delocalization in the obtained tunneling data (extrinsic).

**Quantum tuning of flat band localization**

We further explore the quantum tuning of the flat band localization with a vectorial magnetic field of 1T in Fig. 3(a). It is expected that the magnetic perturbation of the localization will be a third-order tunneling feature in our data, and it should be challenging to detect the corresponding signal. The vector magnetic field has been shown to introduce substantial effects on kagome electrons by scanning tunneling microscopy in several magnetic kagome lattice materials[32], which is beyond the capability of the photoemission technique that works under a zero-field condition. Intriguingly, when we apply magnetic field along one of the three *a*-axes for FeSn, the electronic map at the flat band energy consistently shows a tiny yet detectable intra-unit cell anisotropy (as marked by the colored circles in Fig. 3**a**), which correlates strongly with the magnetic field direction in all three cases. This observation points to a quantum tuning of the flat band localization, which is likely due to the combination of magnetic field tuned vectorial magnetization (spin reorientation) and strong spin-orbit coupling in the system. The effect is not likely from a simple tip effect, as we find that the anisotropy follows the magnetic field direction that we control rather than a random and fixed anisotropy. Moreover, the magnetic field-controlled spin reorientation is consistent with that reported in kagome magnet Mn₃Sn and Fe₃Sn₂ (Refs. 33-36). The zero field and c-axis field data both show much weaker localization anisotropy in real-space, which is likely due to the canting of spins away from the crystalline a-axis reported earlier[22,25]. Since the in-plane magnetic field induced anisotropy is most striking, we try to understand its origin based on a spin-orbit coupled magnetic kagome flat band model, which we discuss below.

In a kagome lattice featuring a spin-polarization along the horizontal direction (Fig. 3**b**), an out-of-plane electric field (from the surface potential) will induce a Rashba spin-orbit coupling, enlarging the hopping of bands not parallel to the horizontal direction. Assume the model without spin-orbit coupling is:

$$H_0 = -t \sum_{<i,j>} c_i^\dagger c_j \qquad (1)$$

where <i,j> stands for the nearest neighbors, $c_i^\dagger$ and $c_j$ are the electron creation and annihilation operators, and $t > 0$ is real. The Rashba spin orbit coupling yields an additional imaginary hopping within each triangular plateau between sublattice sites 1 and 2 (denoted in Fig. 3**b**), as well as between 1 and 3:

$$H_{SOC} = -i\lambda c_1^\dagger c_2 - i\lambda c_1^\dagger c_3 + h.c. \qquad (2)$$

where $\lambda$ is the spin-orbit coupling strength. After a unitary transformation, rotating all the hopping to real, the total Hamiltonian can be written as:

$$H = -\sum_{<i,j>} t_{ij} c_i^\dagger c_j \qquad (3)$$

where $t_{ij} = t$ if $i$ and $j$ belong to sublattices 2 and 3, and $t_{ij} = t' = \sqrt{t^2 + \lambda^2}$ if $i$ and $j$ belong to sublattices 1 and 2 or sublattices 1 and 3. In the momentum space, the Hamiltonian is:

$$H(\mathbf{k}) = \begin{bmatrix} 0 & -2t'\cos\mathbf{k}\cdot\mathbf{a}_{12} & -2t'\cos\mathbf{k}\cdot\mathbf{a}_{13} \\ -2t'\cos\mathbf{k}\cdot\mathbf{a}_{12} & 0 & -2t\cos\mathbf{k}\cdot\mathbf{a}_{23} \\ -2t'\cos\mathbf{k}\cdot\mathbf{a}_{13} & -2t\cos\mathbf{k}\cdot\mathbf{a}_{23} & 0 \end{bmatrix} \qquad (4)$$

under the momentum basis of the 3 sublattices ($|1>_k, |2>_k, |3>_k$), where $\mathbf{a}_{ij}$ is the lattice vector from site $i$ to $j$ ($i, j = 1,2,3$) in Fig. 3**b**. The lowest band is the flat band. We further calculate the charge density $<n_1>$, $<n_2>$, and $<n_3>$ on each sublattice site (that are proportional to the local density of states measured in our experiment) as shown in Fig. 3**c** by setting the system at flat band filling, where $n_i = c_i^\dagger c_i$ is the fermion number operator on site $i$ in Fig. 3**b**. We find $<n_1> > <n_{2,3}>$ when $\lambda \neq 0$, which is consistent with our experimental observations of the geometrical anisotropy of the flat band localization for the kagome lattice layer. Based on this model, we interpret our data as 1) the magnetic field can control the spin direction (magnetization direction) that is consistent with the neutron results[22,25]; 2) through atomic spin-orbit coupling, the spin orientation affects the charge localization anisotropy.

**Conclusion**

Physically, the existence of flat band in the maximally symmetric kagome lattice without spin-orbit coupling is indicated by the presence of localized eigenstates on a hexagon as illustrated in Fig. 3**d**, which has amplitudes with alternating sign, such that any amplitudes outside the hexagon induced by the hopping cancel perfectly. For the Rashba spin-orbit coupled magnetic kagome lattice with horizontal spin-polarization considered here, strictly localized eigenstates no longer exist, since the hoppings along the three directions are no longer equal. However, there still exists one-dimensional eigenstates localized on each horizontal line of bonds parallel to the spin-polarization, with alternating sign of amplitudes as illustrated in Fig. 3**e**. This is because the electron hoppings not parallel to the spin direction are still equal in magnitude and thus cancel with each other. These eigenstates are thus extended (localized) in the parallel (perpendicular) direction to the spin polarization. We expect this magnetization-controlled localization anisotropy physics can be further explored by future transport, spectroscopy and scattering experiments for kagome magnets made into thin films.

**References:**

[1] D. C. Tsui, H. L. Stormer & A. C. Gossard. Two-dimensional magnetotransport in the extreme quantum limit. Phys. Rev. Lett. **48**, 1559–1562 (1982).

[2] R. B. Laughlin. Anomalous quantum Hall effect: an incompressible quantum fluid with fractionally charged excitations. Phys. Rev. Lett. **50**, 1395–1398 (1983).

**Methods**

Single crystals of FeSn are grown out of a Sn flux. 34 g of Sn and 0.33 g of Fe are loaded into a 10-cc alumina crucible and sealed in a silica ampoule under vacuum. The ampoule is heated to 1100 °C, soaked for 12 h, cooled to 1000 °C and soaked for 48 h with occasional shaking of the ampoule, cooled to 800 °C at 6 °C/h, and then cooled to 600 °C at 1 °C/h. Near 600 °C the excess Sn flux is centrifuged into another 10-cc crucible filled with quartz wool.

Single crystals with a size up to 2mm × 2mm are cleaved in situ at 77K in ultra-high vacuum conditions and then inserted into the microscope head, already at $^4$He base temperature (4.2K). Topographic images in this work are taken with the tunneling junction set-up $V = -100$mV and $I = 0.05$nA. Tunneling conductance spectra are obtained with an Ir/Pt tip using standard lock-in amplifier techniques with a lock-in frequency of 1003.3Hz and a junction set-up of $V = -300$mV, $I = 1$nA, and a root mean square oscillation voltage of $V_m = 1$mV. Tunneling conductance maps are obtained by taking dI/dV spectra at each pixel position with a junction set-up of $V = -300$mV, $I = 0.4$nA, and a root mean square oscillation voltage of 10mV.

We estimate the possible renormalization factor for dI/dV spectra taken at different spatial positions. The measured step height of the $Sn_2$ layer is 1.8Å, which is smaller than that in the bulk crystal structure ~2.2Å. One possible reason is that there are variations of the bond distances at the surface, as an extrinsic surface effect. Another possible reason for this difference L = -0.4Å is that the $Fe_3Sn$ layer has much larger local density of states that make the sample-tip height higher even under the same tunneling current and voltage as that in the $Sn_2$ layer, which is noted as an intrinsic effect related to the quantum tunneling principle. The tunneling current is I ~ exp(-2κL), where κ is a work function related coefficient with a typical value of 1Å$^{-1}$ and L is the sample tip distance. Under the assumption that the step height difference is purely caused by the density states difference, then the renormalization factor for the dI/dV spectrum on the $Fe_3Sn$ surface relative to that on the $Sn_2$ surface can be estimated as exp(-2κL) ~ exp(0.8) = 2.2. In a similar spirit, on the $Fe_3Sn$ surface, the measured height between Fe site and Sn site is 0.08Å, while their atomic radius difference is estimated as -0.07Å. Accordingly to this height difference = -0.15Å and under the assumption that there is no surface structure reconstruction, the renormalization factor for the dI/dV spectrum at the Fe site relative to that on the Sn site on the same $Fe_3Sn$ surface can be estimated as exp(-2κL) ~ exp(0.3) = 1.3. We note that while our estimation provides a clue on how Fe atom can feature more density of states near the Fermi level than that on the Sn atom, we have made strong assumptions on the surface structure condition, work function, and atomic radius that would need further experimental confirmation in future.

**Acknowledgement**

M.Z.H. acknowledges support from the US Department of Energy, Office of Science, National Quantum Information Science Research Centers, Quantum Science Center and Princeton University; visiting scientist support at Berkeley Lab (Lawrence Berkeley National Laboratory) during the early phases of this work; support from the Gordon and Betty Moore Foundation (GBMF9461); and support from the US DOE under the Basic Energy Sciences programme (grant number DOE/BES DE-FG-02-05ER46200) for the theory and angle-resolved photoemission spectroscopy work. B.L. is supported by the Alfred P. Sloan Foundation, the National Science Foundation through Princeton University's Materials Research Science and Engineering Center DMR-2011750; and the National Science Foundation under award DMR-2141966. J.-X.Y. acknowledges support from South University of Science and Technology of China principal research grant (number Y01202500). H.L. was supported by National Key R&D Program of China (Grants Nos. 2018YFE0202600 and 2022YFA1403800), Beijing Natural Science Foundation (Grant No. Z200005), National Natural Science Foundation of China (Grants Nos. 12274459). Work at Oak Ridge National Laboratory was sponsored by the U.S. Department of Energy, Office of Science, Basic Energy Sciences, Materials Sciences and Engineering Division. Work at Rice University was supported by US NSF-DMR-2100741 and by the Robert A. Welch Foundation under grant no. C-1839.

**Author contributions**

D.M., J-X.Y., M.S.H. and X.Y. performed the spectroscopic experiment with contributions from H.M., Y-X.J., Y.X., P.D., J.L. and H.D. in consultation with M.Z.H. B.C.S., W.R.M. and H.L. provided the samples. B.L. provided the theoretical analysis in consultation with J-X.Y. and M.Z.H. J.X.Y., B.L. and M.Z.H. wrote the paper and discussed the paper with all authors.

**Competing interests**

The authors declare no competing interests.

**Figures**

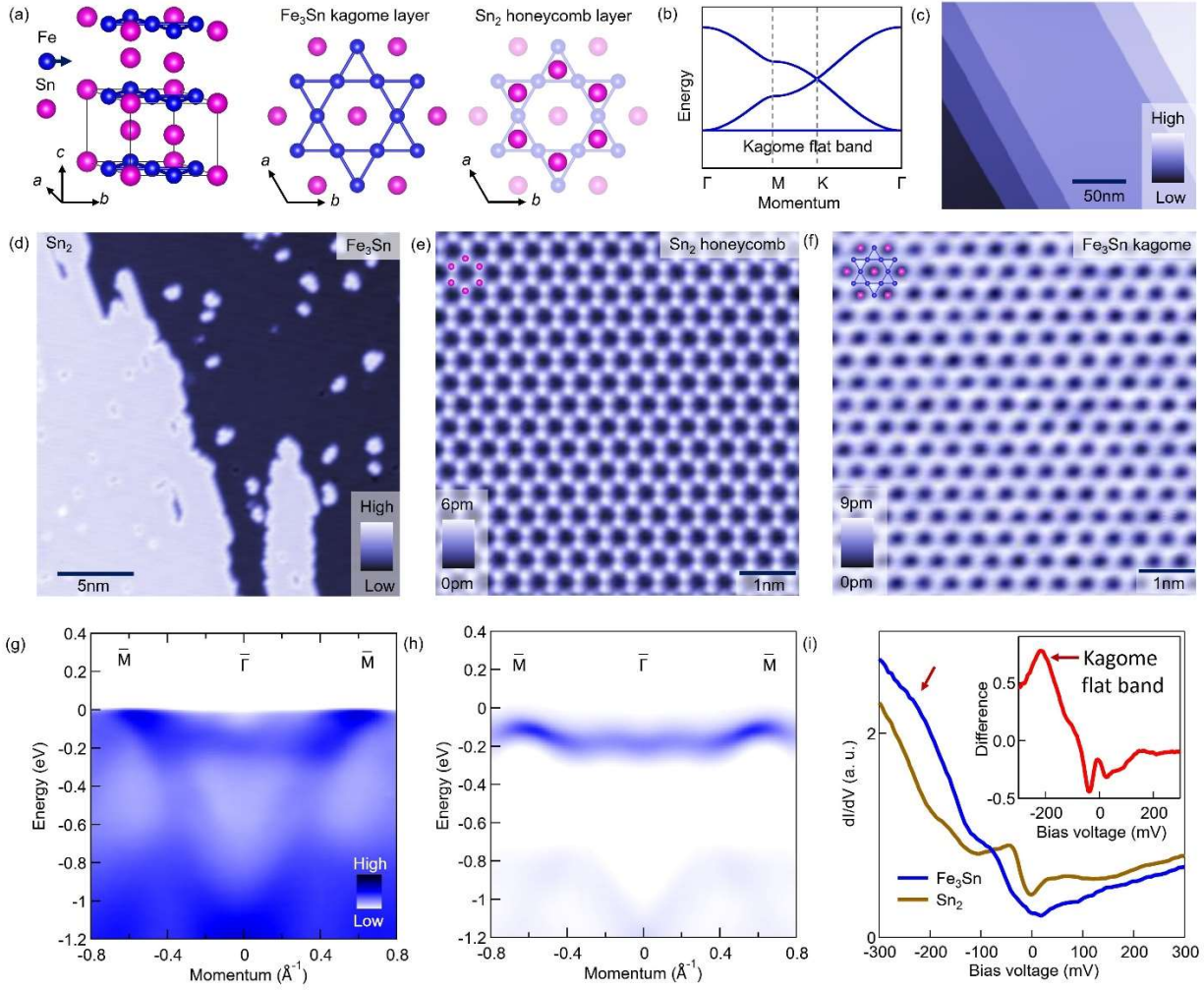

Fig. 1 **Flat band in kagome magnet FeSn.** (a) Crystal structure of FeSn, which is composed of a $Fe_3Sn$ lattice layer and a $Sn_2$ lattice layer. (b) Fundamental band structure based on a tight-bind model in a kagome lattice considering nearest neighboring electron hopping. (c) A topographic image consists of unit-cell steps ($Fe_3Sn$ lattice layers). Data taken with V = -100mV and I = 0.05nA. (d) A topographic image consists of $Fe_3Sn$ and $Sn_2$ lattice layers (V = -100mV, I = 0.05nA). (e) Atomically resolved topographic image of $Sn_2$ honeycomb lattice (V = -100mV, I = 0.1nA). The inset marks the Sn atoms. (f) Atomically resolved topographic image of $Fe_3Sn$ kagome lattice (V = -100mV, I = 0.1nA). The inset marks the Fe and Sn atoms. (g) Angle-resolved photoemission intensity plot for the electron band. Data taken with photon energy of 90eV and temperature of 10K). (h) Symmetrization (along momentum direction) and curvature analysis (along energy direction) of the photoemission data, highlighting the flat band around -0.2eV. (i) Tunneling spectrums taken on $Fe_3Sn$ and $Sn_2$ lattice layers, respectively (V = -300mV, I = 1nA, $V_m$= 1mV). Each spectrum is obtained through averaging spectrums taken over an area of 2nm × 2nm. The inset shows the subtraction of $Fe_3Sn$ spectrum by the $Sn_2$ spectrum, highlighting a peak around -0.2eV as a candidate signature of kagome flat band state.

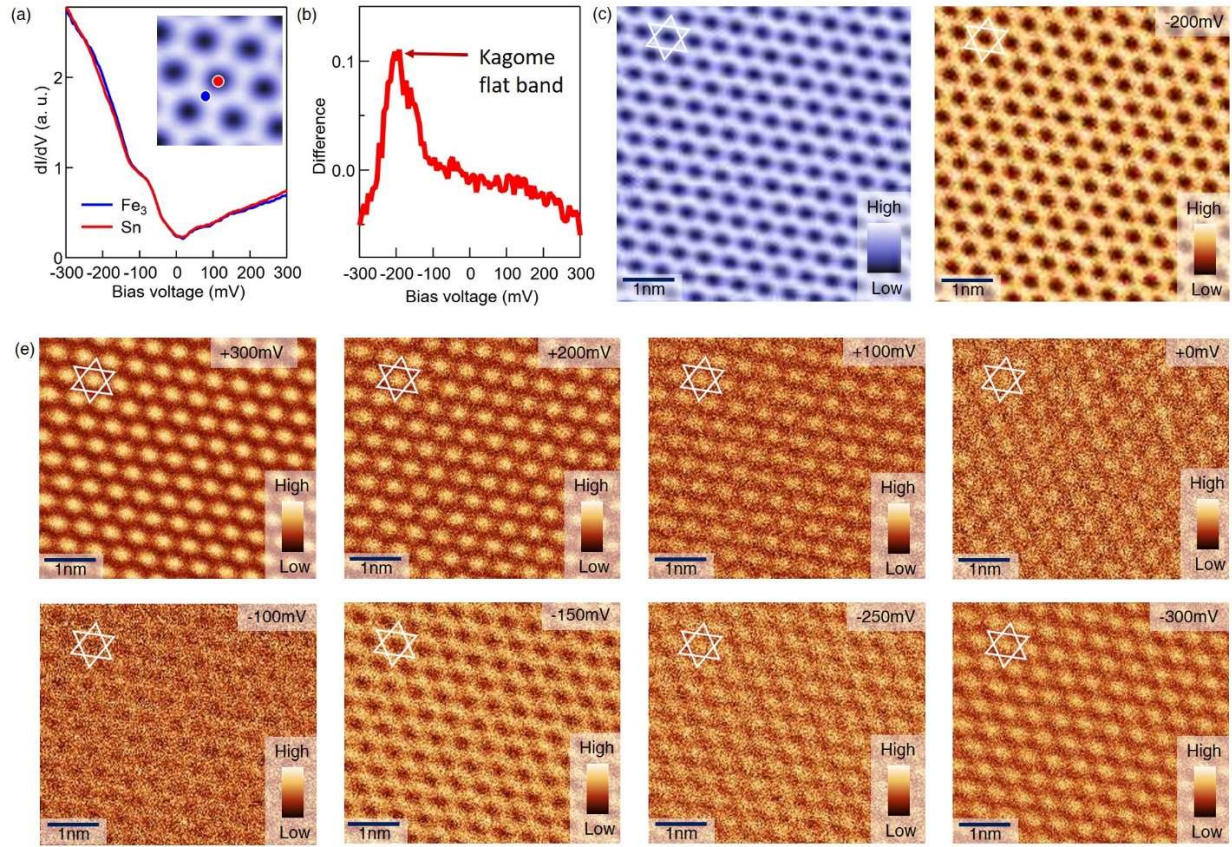

Fig. 2 **Imaging flat band localization.** (a) Tunneling spectrum taken on the Fe and Sn atom sites of the $Fe_3Sn$ kagome layer (as illustrated in the inset), respectively (V = -300mV, I = 1nA, $V_m$= 1mV). (b) The subtraction of the above two spectrums, highlighting a peak around -0.2eV as a candidate signature of kagome flat band state. (c) Topographic image of a $Fe_3Sn$ kagome lattice (V = -300mV, I = 0.4nA). The while lines denote the kagome lattice. (d) Corresponding dI/dV map taken at -0.2eV, showing electronic states localized to the kagome lattice (V = -300mV, I = 0.4nA, $V_m$= 10mV). The while lines denote the kagome lattice. (e) A series of dI/dV maps taken from +0.3eV to -0.3eV, respectively, showing a striking contrast with the -0.2eV map (V = -300mV, I = 0.4nA, $V_m$= 10mV). The while lines denote the kagome lattice for each map.

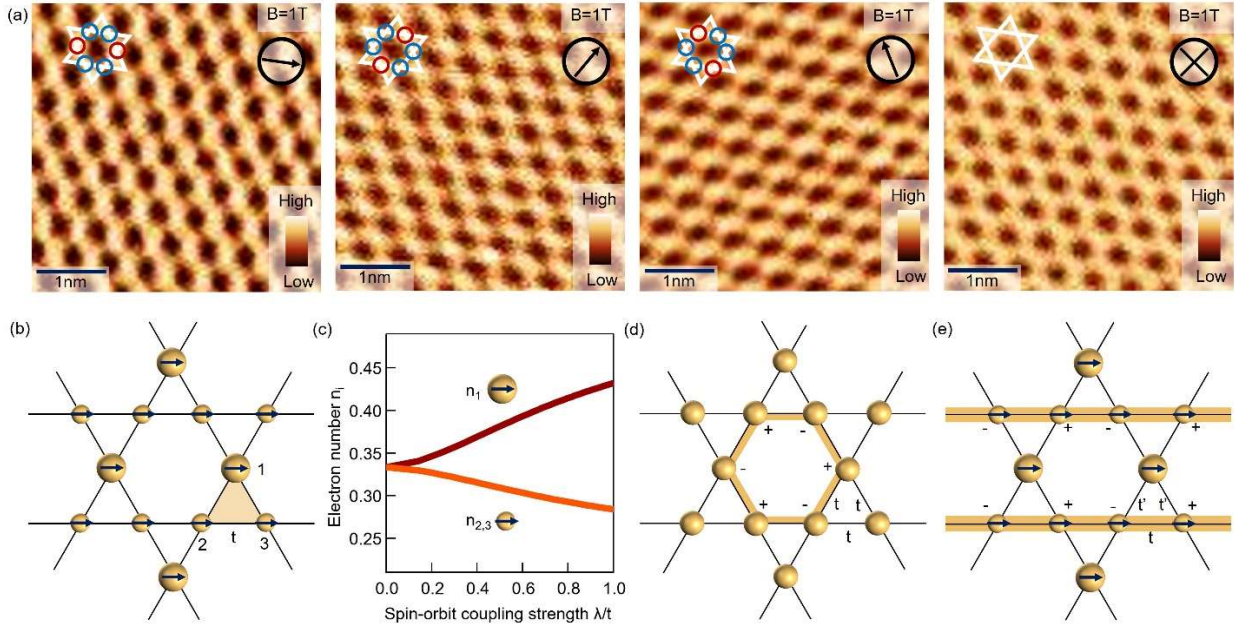

Fig. 3 **Quantum tuning of flat band localization.** (a) dI/dV maps taken at -0.2eV on the $Fe_3Sn$ kagome lattice with in-plane and out-of-plane magnetic fields of 1T (V = -300mV, I = 0.4nA, $V_m$= 10mV). The in-plane magnetic field cause spatial anisotropy of the electronic state within each kagome unit cell (marked by the circles with different colors within each kagome pattern draw by white lines), which correlates strongly with the magnetic field direction. The red circles mark the local intensity related with $\langle n_1 \rangle$, and the blue circles mark the local intensity related with $\langle n_2 \rangle$ or $\langle n_3 \rangle$. (b) Electron number distribution based on a kagome lattice model, which provides an explanation of our experiments. This model considers atomic spin-orbit coupling, ferromagnetic order with spins along the horizontal direction, and flat band electron filling. (c) The charge density anisotropy as a function of spin-orbit coupling strength under the kagome flat band model. (d) The plus and minus signs denote the amplitude of a localized kagome flat band eigenstate. Electron hopping amplitudes outside the golden hexagon are canceled by the destructive quantum interference, leading to the two-dimensional electron localization. (e) In the Rashba spin-orbit coupled magnetic kagome lattice with horizontal spin-polarization, there exist one-dimensional eigenstates localized on each horizontal golden line of bonds, which have alternating signs of amplitude, such that the electron hopping amplitudes perpendicular to the spin direction still cancel with each other.